\documentclass[prd,aps,a4,onecolumn,superscriptaddress,preprintnumbers,nofootinbib,notitlepage]{revtex4-1}

\usepackage{pslatex}
\usepackage[pdftex]{graphicx}
\usepackage{psfrag}
\usepackage{epsfig}
\usepackage{color}
\usepackage{cancel}
\usepackage{slashed}
\usepackage{amssymb}
\usepackage{amsmath}
\usepackage{hyperref}
\usepackage{enumerate}
\usepackage{multirow}
\usepackage{ulem}
\usepackage{float}
\usepackage{comment}
\usepackage{diagbox}
\usepackage{amsmath}

\usepackage{multirow}
\usepackage{tabularx}
\usepackage{graphicx}
\usepackage{subfigure}
\usepackage{caption}
\usepackage{mathrsfs}

\begin{document}

\preprint{MI-HET-812}


\title{Solar neutrinos with CE$\nu$NS and flavor-dependent radiative corrections}

\author{Nityasa Mishra}
\email{nityasa_mishra@tamu.edu}%
\affiliation{Department of Physics and Astronomy, Mitchell Institute for Fundamental Physics and Astronomy, Texas A\&M University, College Station, Texas 77843, USA}
\author{Louis E. Strigari}%
\email{strigari@tamu.edu}%
\affiliation{Department of Physics and Astronomy, Mitchell Institute for Fundamental Physics and Astronomy, Texas A\&M University, College Station, Texas 77843, USA}

\date{\today}

\begin{abstract}
We examine solar neutrinos in dark matter detectors including the effects of flavor-dependent radiative corrections to the CE$\nu$NS cross section. Working within a full three-flavor framework, and including matter effects within the Sun and Earth, detectors with thresholds $\lesssim 1$ keV and exposures of $\sim 100$ ton-year could identify contributions to the cross section beyond tree level. The differences between the cross sections for the flavors, combined with the difference in fluxes, would provide a new and unique method to study the muon and tau components of the solar neutrino flux. Flavor-dependent corrections induce a small day-night asymmetry of $< |3 \times10^{-4}|$ in the event rate, which if ultimately accessible would provide a novel probe of flavor oscillations. 
\end{abstract}

\maketitle

\section{Introduction \label{sec:introduction}}

\par The cross section for coherent elastic neutrino-nucleus scattering (CE$\nu$NS) in the Standard Model (SM) is theoretically well-understood at tree-level~\cite{Abdullah:2022zue}. Because this is a neutral current process, at tree-level the cross section is the same for all neutrino flavors. Using a stopped-pion beam, which produces electron, muon, and anti-muon neutrinos with energies $\mathcal{O} (10)$ MeV, COHERENT has measured the CE$\nu$NS cross section on Ar and CsI targets. Summed over neutrino flavors produced by the source, this measurement is consistent with the SM prediction~\cite{COHERENT:2021xmm}. 

\par Forthcoming COHERENT data is expected to improve on this measurement for different nuclear targets~\cite{Akimov:2022oyb}. In addition, reactor experiments, which are sensitive to electron anti-neutrinos, will measure the CE$\nu$NS cross section at lower neutrino energies, $\mathcal{O} (1)$ MeV. Solar neutrinos, which will be detected via CE$\nu$NS at future dark matter experiments, will measure the cross section for neutrino energies roughly between stopped pion and reactor experiments. Interestingly, given the flavor composition of the solar flux that arrives at Earth, dark matter experiments will be sensitive to electron, muon, and tau neutrinos via the CE$\nu$NS process. 

\par Beyond tree level in the SM, it has long been known that the CE$\nu$NS cross section is flavor-dependent~\cite{Sehgal:1985iu}. The simplest way to see this is through the neutrino charge radius diagram, which in the limit of small momentum transfer amounts to a lepton-mass dependent shift in the $\sin^2 \theta_w$ term in the cross section~\cite{Papavassiliou:2005cs,Cadeddu:2018dux,Tomalak:2020zfh}. The flavor-dependent corrections scale as the mass of the charged lepton, implying that they are the most significant for tau neutrinos. In addition to flavor-dependent corrections, there are flavor-independent corrections that affect electron, muon, and tau neutrinos in a similar manner. Radiative corrections may be detectable in future stopped-pion experiments with reduced systematic uncertainties~\cite{Tomalak:2020zfh}. 

\par In this paper, we study the impact of SM radiative corrections on the detection of solar neutrinos via CE$\nu$NS at future direct dark matter detection experiments. Solar neutrinos are a unique source for CE$\nu$NS because all three flavor components are present in the flux. In order to determine the interaction rate of electron, muon, and tau neutrinos, we account for matter effects in both the Earth and the Sun within a full three-flavor neutrino oscillation framework. The full three-flavor analysis is necessary, since the cross section is different for each of the three flavors when including radiative corrections. 

\par We highlight two particular interesting phenomenological implications of radiative corrections in solar neutrino CE$\nu$NS measurements. The first is a possible separation of the muon and tau neutrino flavor components of solar flux. Previous solar neutrino experiments~\cite{SNO:2011hxd,Super-Kamiokande:2016yck} with neutral current sensitivity utilized cross sections that are the same for muon and tau neutrino flavors. Therefore through precise CE$\nu$NS measurements, dark matter experiments could differentiate between muon and tau flavors by exploiting both the differences in their fluxes and cross sections. The second interesting phenomenological implication is matter-induced oscillation effects, which introduce a day-night asymmetry in the propagation of electron neutrinos in the CE$\nu$NS channel. This effect has been measured by Super-Kamoikande using electron-neutrino scattering, but has yet to be studied using a different neutrino detection channel. We provide the first estimate of the day-night asymmetry for neutrinos via a channel different from neutrino-electron elastic scattering. 

\par This paper is organized as follows. In Section~\ref{sec:crosssection} we review the CE$\nu$NS cross section with radiative corrections implemented. In Section~\ref{sec:matter} we discuss how matter effects, both those in the Sun and Earth, affect neutrino propagation within a three-neutrino framework. In Section~\ref{sec:results} we present the results of our analysis, and in Section~\ref{sec:discussion} our discussion and conclusions. 

\section{CE$\nu$NS with radiative corrections}
\label{sec:crosssection} 
The tree level CE$\nu$NS differential cross section as a function of neutrino energy $E_\nu$, recoil energy $T$, and nuclear mass $M_A$ is~\cite{PhysRevD.97.033003}
\begin{align}
     \frac{d \sigma_{\nu}}{dT} = \frac{G_F^2 M_A}{\pi}(Q^V_W)^2 \left[1 - \frac{T}{E_\nu} - \frac{M_A T}{2 E_\nu^2} \right] F^2_W(Q^2) +  \frac{G_F^2 M_A}{\pi}(Q^A_W)^2 \left[1 - \frac{T}{E_\nu} +
       \frac{M_A T}{2 E_\nu^2}\right]  F^2_W(Q^2)  
    \label{eq:tree cross-section with axial}   
\end{align}
where $G_F$ is the Fermi coupling constant and the nuclear mass is $M_A = A \times 931$ MeV where the mass number $A=Z+N$ with proton number $Z$ and neutron number $N$. The 4-momentum transfer $Q^2$ is related to nuclear recoil as $T = Q^2/2M_A$, and takes values in the range [0, $T_{max}$], where
\begin{equation}
    T_{max} = \frac{2 E^2_\nu}{M_A + 2E_\nu}. 
\end{equation}
The factors $Q^V_W$and $Q^A_W$ refer to vector and axial vector weak charges respectively. The vector charge is  
\begin{equation}
Q^V_W = g_p^VZ + g_n^VN
\label{eq:QVW} 
\end{equation} 
where $g_p^V = 1/2 - 2\sin^2{\theta_w}$ and $g_n^V=-1/2$ are the proton and neutron vector coupling to $Z^0$.   
The axial charge is 
\begin{equation} 
 Q^A_W = g_p^A(Z_+ - Z_-) + g_n^A(N_+ - N_-)
 \end{equation} 
 where $Z_\pm(N_\pm)$ refer to the spin up (+) and spin down (-) protons (neutrons), and $g_p^A = 0.63$ and $g_n^A = -0.59$ are the axial vector couplings with the $Z^0$. For spin-zero nuclei, $Q_W^A =0$, and for nuclei with spin, $\frac{Q_W^A}{Q_W^V}\sim \frac{1}{A}$. As a result the axial contribution term in the cross section is strongly suppressed relative to the vector contribution. 

Thus for the tree level differential cross-section we can write 
\begin{equation}
     \frac{d \sigma_{\nu}}{dT} = \frac{G_F^2 M_A}{4\pi} \left(1 - \frac{T}{E_\nu} - \frac{M_A T}{2 E_\nu^2} \right) Q_W^2F^2_W(Q^2) 
    \label{eq:tree cross-section}  
\end{equation}
where the weak charge at tree-level is given by $Q_W = N - (1-4 \sin{\theta_w}^2 )Z$ with $\sin^2\theta_w$ as the weak mixing angle. Note that $Q^V_W$ differs from $Q_W$ by a factor of two. 
The $F_W$ term in Equation~\ref{eq:tree cross-section with axial} represents the weak form factor, which encodes the information on the distribution of the nucleons in the nucleus. This can be approximated as 
\begin{equation*}
    F_W = \frac{1}{Q_W}[N F_n(Q^2) - (1-4\sin^2{\theta_w})Z F_p(Q^2)]
\end{equation*}
where we take the Helm form factor prescription~\cite{LEWIN199687} for $F_n$ and $F_p$ representing neutron and proton form factors, respectively (see also Ref.~\cite{AristizabalSierra:2019zmy} for analysis of COHERENT data and implications for form factor models). 


Including next-to-leading order effects change Equation~\ref{eq:tree cross-section}. In the Effective Field Theory (EFT) formalism, the next-to-leading order term in the cross section is~\cite{Tomalak:2020zfh}  
\begin{equation}
    \frac{d \sigma_{\nu l}}{dT} = \frac{G_F^2 M_A}{4 \pi} \left(1 - \frac{T}{E_\nu} - \frac{M_A T}{2 E_\nu^2} \right) {\cal F}_{\nu l}^2 (Q^2)
     \label{eq:rad cross-section} 
\end{equation}
where 
\begin{equation}
    {\cal F}_{\nu l}(Q^2) = {\cal F}_W(Q^2) + \frac{\alpha}{\pi}[\delta^{\nu l} + \delta^{QCD}] {\cal F}_{ch}(Q^2), 
     \label{eq:rad form factor} 
\end{equation}
with ${\cal F}_W(Q^2)$ and ${\cal F}_{ch}(Q^2)$ defined as the re-normalization-scale dependent weak and charge form factors. The factors $\delta^{\nu l}$ represent flavor-dependent corrections that arise from charged current loop corrections while $\delta^{QCD}$ represents the corrections due to quark and hadronic loops~\cite{Tomalak:2020zfh}. 

It is important to note here that ${\cal F}_W(Q^2)$ is different from $F_W(Q^2)$. This can be understood by ``turning off" the radiative corrections associated with $\alpha$, i.e. $\alpha\rightarrow 0$. In this case, Equation~\eqref{eq:rad cross-section} still differs from the tree-level result in Equation~\eqref{eq:tree cross-section}. In other words, there are radiative corrections to the weak form factor ($F_W$) and the weak-charge ($Q_W$) that are embedded in the term ${\cal F}_{\nu l}$. 

\begin{figure*}[h]
    \centering
    \includegraphics[width=0.51\textwidth]{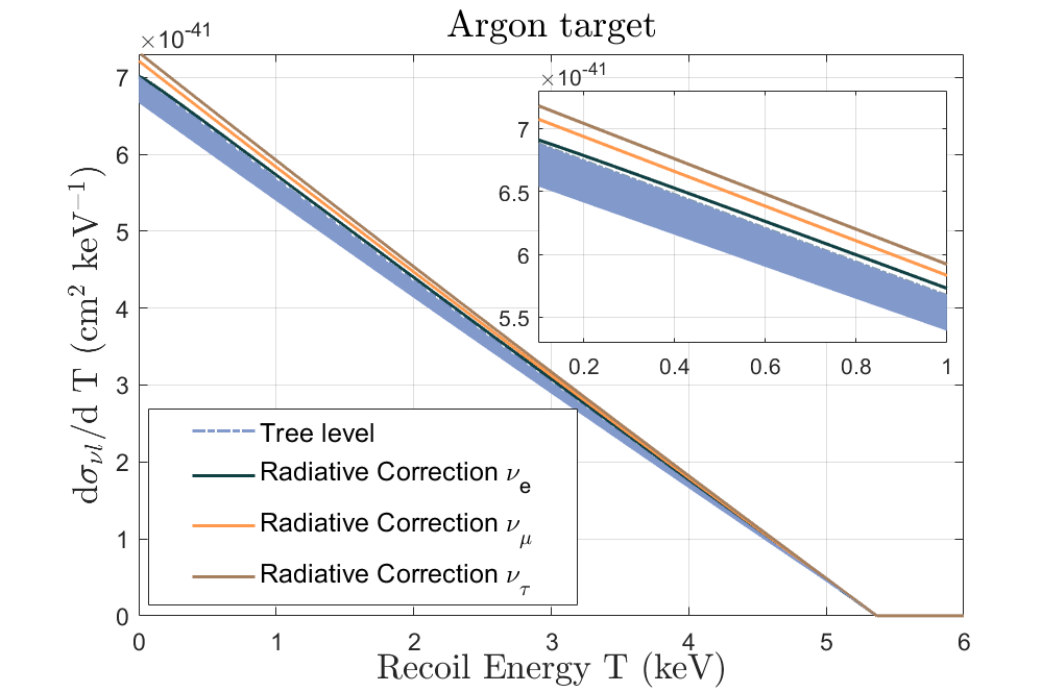}
    \hspace*{-0.6cm}
    \includegraphics[width=0.51\textwidth]{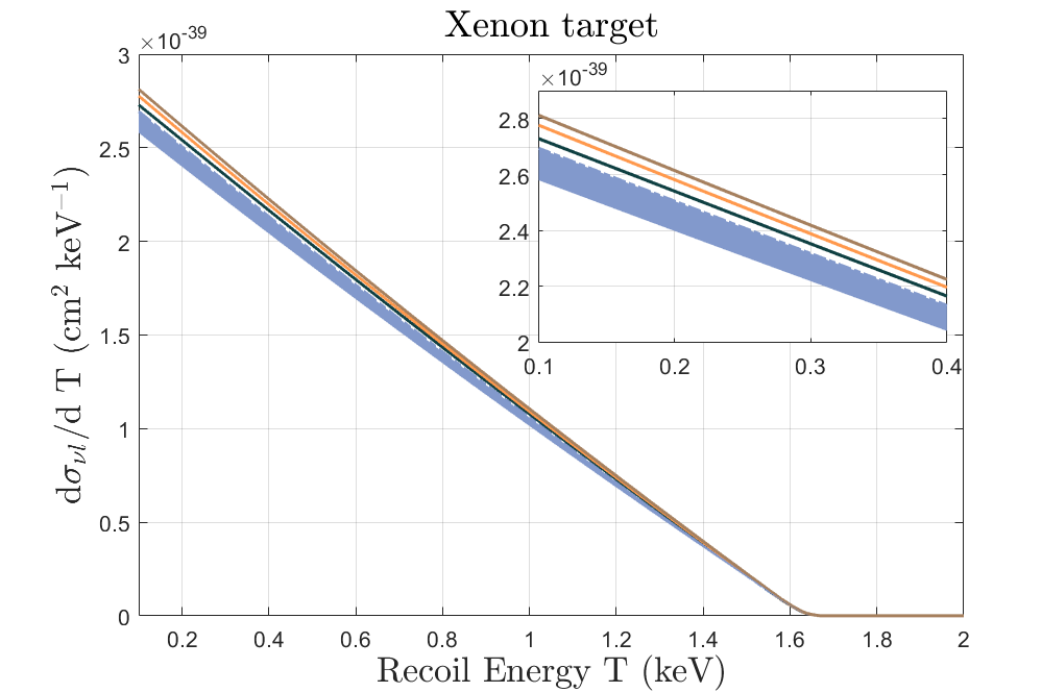}
    \caption{Cross section for a 10 MeV neutrino for an Argon (left) and Xenon (right) target. The upper curves are the radiative correction models for each of the different flavors ($l$), as indicated in the legend, while the lower band shows the tree level calculation for the range of allowed $\sin^2 \theta_w$. The insets zoom in on the regions at low recoil energies where the deviations between the cross sections become most apparent. }
    \label{fig:diff_cross}
\end{figure*}

\par In Figure~\ref{fig:diff_cross}, we show the cross section as a function of nuclear recoil energy for neutrinos with energy 10 MeV, for Xenon and Argon targets. Here, we see that the cross sections, which are different for electron, muon, and tau flavors, start to diverge particularly at the lowest recoil energies. For comparison, we show the range of tree-level predictions for different assumed values of $\sin^2 \theta_w$, with a value of $\sin^2 \theta_w = 0.23858$ motivated from Table 10.2 of Ref.~\cite{PhysRevD.98.030001}, and $\sin^2 \theta_w = 0.23112$ as motivated from previous CE$\nu$NS analyses~\cite{Tomalak:2020zfh}. Even accounting for this range of $\sin^2 \theta_w$, the flavor-corrections are shown to separate from the range of tree level predictions. For example, at $T = 1$ keV and $E_\nu = 10$ MeV, and for $\sin^2 \theta_w = 0.23858$ the percent difference between the tree level model and the radiative correction is 0.95$\%$ (1.8$\%$) for $\nu_e$, 2.77$\%$ (3$\%$) for $\nu_\mu$ and 4.32$\%$ (4.4$\%$) for $\nu_\tau$ for Argon (Xenon). Therefore, an experiment with sufficient sensitivity to low-energy recoils should be able to distinguish the tree level from radiative correction models. 

\section{Solar neutrinos and Earth-matter effects} 
\label{sec:matter} 
In this section we briefly review matter effects in the propagation of solar neutrinos, highlighting how they are relevant for our calculations. We separately account for both matter effects in the Sun and in the Earth within a full three-flavor framework.


\subsection{Propagation of solar neutrinos in the sun}

At the core of the Sun, electron neutrinos ($\nu_e$) are produced primarily via the proton-proton (pp) chain, with a small component produced from the Carbon-Nitrogen-Oxygen (CNO) cycle~\cite{Haxton:2012wfz}. To determine the effect of flavor transformations on neutrinos as the propagate through the Sun, we consider a standard three-flavor model, in which the neutrino propagation is adiabatic, so that the density variation is imperceptible or ``slow" along an oscillation length.

\begin{table*}[htbp]
\centering
\begin{tabular}{|p{4cm}<{\centering}|p{4cm}<{\centering}|p{4cm}<{\centering}|p{4cm}<{\centering}|}\toprule
\textbf{Reaction} & $\mathbf{E_\nu}$\textbf{(MeV)} & \multicolumn{2}{|c|}{\textbf{Maximum Recoil Energy (keV)}} \\
\hline
& & \textbf{Argon Target} & \textbf{Xenon Target}\\
\hline
pp & $<$ 0.42 & 9.47 x $10^{-3}$ & 2.96 x $10^{-3}$ \\
\hline
\multirow{2}{*}{$^7$Be} & 0.38 & 7.92 x $10^{-3}$ & 2.48 x $10^{-3}$ \\
 & 0.86 & 3.99 x $10^{-2}$ & 1.24 x $10^{-2}$ \\
\hline
CNO-N & $<$ 1.20 & 7.72 x $10^{-2}$ & 2.41 x $10^{-2}$ \\
\hline
pep & 1.44 & 0.112 & 3.49 x $10^{-2}$ \\
\hline
CNO-O & $<$ 1.73 & 0.161 & 5.02 x $10^{-2}$ \\
\hline
CNO-F & $<$ 1.74 & 0.163  & 5.08 x $10^{-2}$ \\
\hline
$^8$B & $<$ 16.35 & 14.34  & 4.49 \\
\hline
hep & $<$ 18.77 & 18.90  & 5.91  \\
\hline
\end{tabular}
\caption{Solar neutrino sources and their energy ranges, along with the maximum recoil energy the source component can impart. If the maximum recoil energy is below the threshold of the detector, the associated component will not contribute to the event rate.}
\label{tab:Solar neutrino flux}
\end{table*}

As the neutrinos travel through the Sun, the electron number density $N_e$ changes. This induces a corresponding variation in the matter potential, which is given by 
\begin{align}
    V_{cc} = \sqrt{2} G_F N_e. 
    \label{eq:Vcc}
\end{align}
The electron density profile of the Sun at a given time of propagation ($t$) can be modeled as~\cite{PETCOV1988139}, 
\begin{equation}
    N_e(t) 
    = N_{e}(t_0)\exp \left( - {\frac{10 x}{R_\odot}} \right) 
\end{equation}
where $x \sim t-t_0$ is the distance traveled by the neutrino in the Sun since production time $t_0$, and  $ R_\odot \simeq 7 \times 10^8$ m represents the solar radius. The electron number density decreases almost monotonically from 100$N_{A} \,\,\, \textrm{cm}^{-3}$  at the center to zero at the surface of the Sun. As a result, the adiabatic parameter for neutrino of energy $E_\nu$, given by $\gamma = |\frac{-\Dot{V}}{V}\frac{\cos2\theta E_\nu}{ \Delta m^2\sin^2 2\theta}|$, is small leading to an incoherent neutrino flux of the mass-eigenstates $|\nu_{i}\rangle = (\nu_1, \nu_2, \nu_3)^T$ as neutrinos exit the Sun. Here $\theta$ represents the mixing angle whereas $\Delta m$ is the mass-splitting, as for the case of standard neutrino oscillation. 

\par The probability of an electron neutrino transitioning to a flavor $\alpha$ at the surface of the Sun is given by
\begin{align}
    P^S_{e\alpha} = P^S_{e1}P_{1\alpha} + P^S_{e2}P_{2\alpha} + P^S_{e3}P_{3\alpha}. 
    \label{eq: Prob sun}
\end{align}
Here $P^S_{ei}$ refers to the adiabatic transition probability of a $\nu_e$ to a $\nu_i$ as the neutrino travels through the Sun, and $P_{i\alpha}$ provides information on the fraction of $\nu_\alpha$ present in $\nu_i$, i.e. $P_{i\alpha} = |\langle \nu_\alpha | \nu_i\rangle|^2$ , where $|\nu_{\alpha}\rangle = (\nu_e, \nu_\mu, \nu_\tau)^T$ are the flavor states. The relation between the mass eigenstates and the flavor eigenstates is  
\begin{align}
    |\nu_{i}\rangle = \sum_\alpha U_{\alpha i}|\nu_{\alpha}\rangle. 
\label{eq: nu_i U nu_fl}
\end{align}
We use the standard parametrization of the unitary matrix given by the PMNS matrix
\begin{align*}
    U &= \begin{bmatrix}
       1 & 0 & 0 \\
       0 & c_{23} & s_{23} \\
       0 & -s_{23} & c_{23}
        \end{bmatrix} 
        \begin{bmatrix}
        c_{13} & 0 & s_{13}e^{-i\delta_{CP}} \\
        0 & 1 & 0\\
        -s_{13}e^{i\delta_{CP}} & 0 & c_{23}
        \end{bmatrix}
        \begin{bmatrix}
        c_{12} & s_{12} & 0  \\
        -s_{12} & c_{12} & 0 \\
        0 & 0 & 1
        \end{bmatrix} 
        = \begin{bmatrix}
       c_{12}c_{13} &  s_{12}c_{13} & s_{13}e^{-i\delta_{CP}}\\
       -s_{12}c_{23} -c_{12}s_{23}s_{13}e^{i\delta_{CP}} & c_{12}c_{23} -s_{12}s_{23}s_{13}e^{i\delta_{CP}} & s_{23}c_{13}\\
        s_{12}s_{23} -c_{12}c_{23}s_{13}e^{i\delta_{CP}} & -c_{12}s_{23} -s_{12}c_{23}s_{13}e^{i\delta_{CP}} & c_{23}c_{13}
        \end{bmatrix} 
\end{align*}
where $c_{ij} = \cos \theta_{ij}$ and $s_{ij} = \sin \theta_{ij}$ with $\theta_{ij}$ refers to the mixing angles and $\delta_{CP}$ is the CP-violating phase. For the parameters describing the PMNS matrix, we use the best fit values for these parameters~\cite{nu_fit}: 
\begin{align*}
    &\text{mixing angles:} && \theta_{12} = 33.45^\circ && \theta_{23} = 49.2^\circ && \theta_{13} = 8.57^\circ \\
    &\text{ mass-splittings:} && \Delta m_{21}^2 = 7.5 \times 10^{-5} \text{eV}^2 && \Delta m_{3l}^2 = 2.5 \times 10^{-3} \text{eV}^2
\end{align*}
and take $\delta_{CP} = 135^\circ$ which is the maximal CP-violating value and is consistent with the allowed 3-sigma range.

The Hamiltonian describing propagation in the Sun is made up of the local eigenvalues and can be written as
\begin{align}
     H_{fl}|\nu_{i}(t)\rangle = \hat{U}(t) E_M(t)\hat{U}^\dagger(t)|\nu_{\alpha}(t)\rangle
     \label{eq:Scrondingers eq sun}
\end{align}
where the unitary mixing matrix $\hat{U}$ is written in terms of the local matter mixing angles $\theta_m$, the local flavor eigenstates, $\nu_{\alpha}(t)$, and mass eigenstates, $\nu_{i}(t)$, that depend on the position and/or the time along the path of neutrino transit. The energy eigenvalue matrix in the mass basis is $E_M = \frac{1}{2E_\nu} \textrm{diag}(m_1^2,m_2^2,m_3^2)$. Consequently the $P^S_{ei}$ can be written as~\cite{Blennow_2004}
\begin{align*}
    P^S_{ei} &= \int_0^{R_\odot}|\langle \nu_i|\nu_e(t)\rangle|^2f(r)dr =\int_0^{R_\odot}\sum_{j=1}^3|\hat{U}_{ej}(\theta_m(N_e(r))|^2 P^{jump}_{ji}f(r)dr
\end{align*}
where $ P^{jump}_{ji}$ refers to the probability of jumping from one energy eigenvector to the other. For adiabatic propagation, $P^{jump}_{ji}= \delta_{ji}$.

Neutrinos from the different nuclear reactions are produced within different regions of the Sun. To account for this, we average (represented by $\langle\Bar{f}\rangle$ for average of any function $f$) over the different production zones within the Sun, weighted by the fraction of neutrinos produced in each zone $f(r)$. The information on zonal neutrino production fraction and flux information is taken from Ref.~\cite{Bahcall_rep}.

For mixing between three-flavors of neutrinos, there are two possible resonances, one for each mass-splitting given by $V_{cc}^{res} \simeq \frac{\Delta m^2}{2 E_\nu} \cos\theta$. The $V^{sun}_{cc}$, which is a function of the electron number density in the Sun as in Equation~\ref{eq:Vcc}, is much smaller than the $V_{cc}^{res}$ for $\Delta m_{3l}^2$ i.e $V_{cc}^{sun} = \sqrt{2} G_F N^{sun}_e<<V_{cc_{max}}^{res} = \frac{\Delta m^2_{3l}}{2 E_\nu}$ . See reference~\cite{Blennow_2004} for a thorough discussion. As a result one can thus consider that $\nu_3$ ``decouples" from the other two states giving an effective three-flavor probability
\begin{align}
   &P^S_{e1} = \frac{c_{13}^2}{2}[1+\langle \overline{\cos{2\theta_{{12}_m}}}\rangle] \nonumber\\
   &P^S_{e2} = \frac{c_{13}^2}{2}[1-\langle \overline{\cos{2\theta_{{12}_m}}}\rangle] \nonumber\\
   &P^S_{e3} \simeq \sin^2 {\theta_{13}}
   \label{eq: Pei sun}
\end{align}
where $ \cos2\theta_{{12}_m} = (k \cos 2\theta_{12} - V^{eff}_{CC})/k_m $ and $k_m = \sqrt{(k\cos2\theta_{12} - V^{eff}_{CC})^2 + (k\sin2\theta_{12})}$. The term $k = \frac{\Delta m^2}{2E_\nu}$. The effective potential is given by $V^{eff}_{cc} = \cos^2 \theta_{13} V_{cc}$~\cite{Blennow_2004}.

\subsection{Propagation through Earth}
We now move on to propagation in the Earth. Similar to Equation~\ref{eq: Prob sun}, the probability after transitioning the Earth matter, $P^E$, is given by 
\begin{align}
    P^E_{e\alpha} = P^S_{e1}P^E_{1\alpha} + P^S_{e2}P^E_{2\alpha} + P^S_{e3}P^E_{3\alpha}
    \label{eq: Prob earth}
\end{align}
where $P^E_{i\alpha}$ represents the probability of transition from mass state $\nu_i$ to $\nu_\alpha$ along the neutrino path through the Earth. 

In order to calculate $P^E_{i\alpha}$ we consider the Hamiltonian describing the evolution of mass-eigenstates $|\nu_i\rangle$
\begin{align}
      H_{M}|\nu_{i}\rangle = (E_M + U^\dagger VU)|\nu_{i}\rangle.
     \label{eq:Scrondingers eq earth}
\end{align}
 Here $V$ = diag$(V_{cc}(t), 0, 0)$ is the matter potential. The non-zero element in the matrix $V$  for the electron neutrino arises because the charged current interaction is only viable for $\nu_e$ from the presence of electrons in matter. While neutral current interactions do occur, they can be subtracted from the matter potential as they do not affect the oscillations.

\par
\begin{figure*}[h]
    \centering
    \hspace*{-1.cm}
    \includegraphics[width=1.1\textwidth]{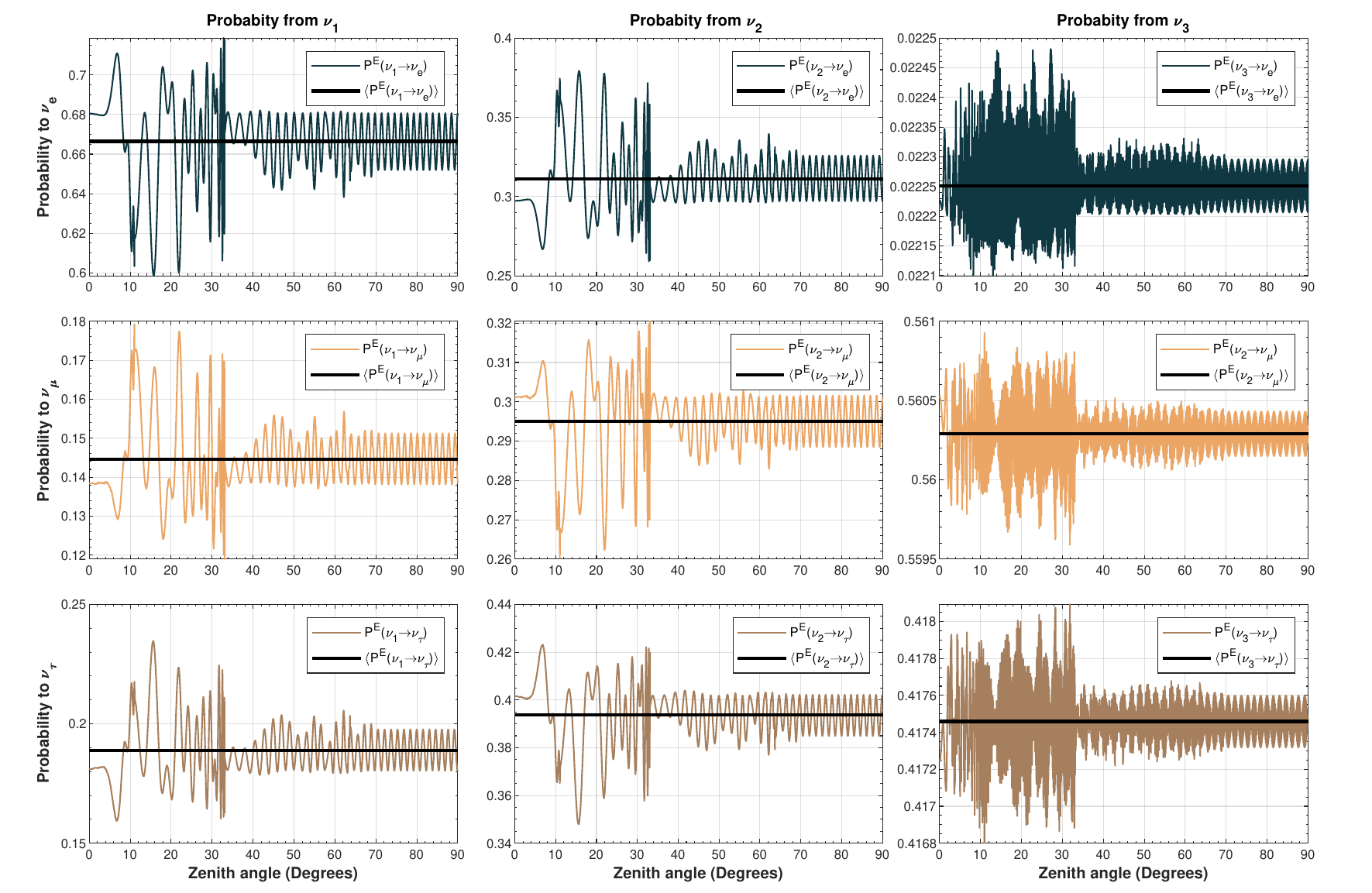}
    \caption{Probabilities for neutrinos propagating through the Earth as a function of zenith angle, in degrees, for a neutrino of energy 10 MeV. The different rows represent probabilities for transitions to the three flavors, while the different columns represent transitions from different mass eigenstates.} ~\label{fig:zenith} 
\end{figure*}

 The amplitude of transition is $A_{i\alpha} = \langle\nu_\alpha|\nu_i(t)\rangle$ with $\langle\nu_{\alpha}| =\sum_i\langle\nu_i|U_{\alpha i}$. The transition probabilities are then 
\begin{align}
    P^E_{i\alpha} &= |\langle\nu_\alpha|\nu_i(t)\rangle|^2 =|\sum_{j,k}\langle\nu_j|U_{\alpha j}S_{ik}|\nu_k(0)\rangle|^2 
    = |\sum_{j}U_{\alpha j}S_{ij}|^2 = |[S U^T]_{i\alpha}|^2
    \label{eq: Pie earth}
\end{align}
where $S$ is the transition matrix in mass-basis, given by $\exp{(-iH_M L)}$, with $L$ being the distance travelled by the neutrino. Combining Equations~\eqref{eq: Pei sun} and~\eqref{eq: Pie earth} with Equation~\eqref{eq: Prob earth}, we obtain the probability of detecting neutrinos of a particular flavor in the night after traveling through Earth (for comparison, the probability of detecting electron neutrinos in the day is given by $P^S_{e\alpha}$ in Equation~\eqref{eq: Prob sun}, which is the probability of detecting $\nu_\alpha$ as it exits the surface of the Sun). The resulting probabilities $P^E_{i\alpha}$ are shown in Figure~\ref{fig:zenith}, for the example case of a $10$ MeV neutrino. Here we model the Earth density profile using the five-layer model~\cite{DZIEWONSKI1981297} of constant matter density solutions for the matter effect. The change in amplitude of the probabilities that are evident result from transitions through different layers in the Earth. 

\subsection{Averaging of survival probability}

Since the detectors that we consider do not have sensitivity to the direction of the incoming neutrino, we must average over the incoming neutrino direction  as well as time. We then average the  probabilities in Equation~\ref{eq: Pie earth} as 

~\cite{Lisi_1997};
\begin{align}
    \langle P^E\rangle = \frac{\int^{\tau_{d_2}}_{\tau_{d_1}} d\tau_d\int^{\tau_{h_2}(\tau_d)}_{\tau_{h_1}(\tau_d)} d\tau_h P^E(\eta(\tau_h,\tau_h))}{\int^{\tau_{d_2}}_{\tau_{d_1}} d\tau_d\int^{\tau_{h_2}(\tau_d)}_{\tau_{h_1}(\tau_d)} d\tau_h}
\end{align}
 where $\eta$ is the nadir angle ($180^\circ$ - zenith angle) of the Sun at the detector site. Since the nadir angle will change based on the Sun position in the sky during the day (and also the night), the nadir angle is a function of daily time $\tau_d$ and hourly time $\tau_h$. The above equation can then be written as a single integral of the form
\begin{align}
   \langle P^E\rangle = \int^{\eta_2}_{\eta_1}d\eta W(\eta) P^E(\eta)
   \label{eq:pe} 
\end{align}
Here $W(\eta)$ can be thought of as a weight function that gives information of ``solar exposure" of the trajectory. For more details on the weight function see Ref.~\cite{Lisi_1997}. For our analysis we use a detector latitude of $\lambda = 42.421^\circ$ corresponding to the latitude for  Gran Sasso National Laboratory. However, on repeating the calculation for other latitudes, we found that the results  do not change appreciably for this analysis.

\par
\begin{figure*}[h]
    \hspace*{-1.5cm}
    \includegraphics[width =1.2\textwidth]{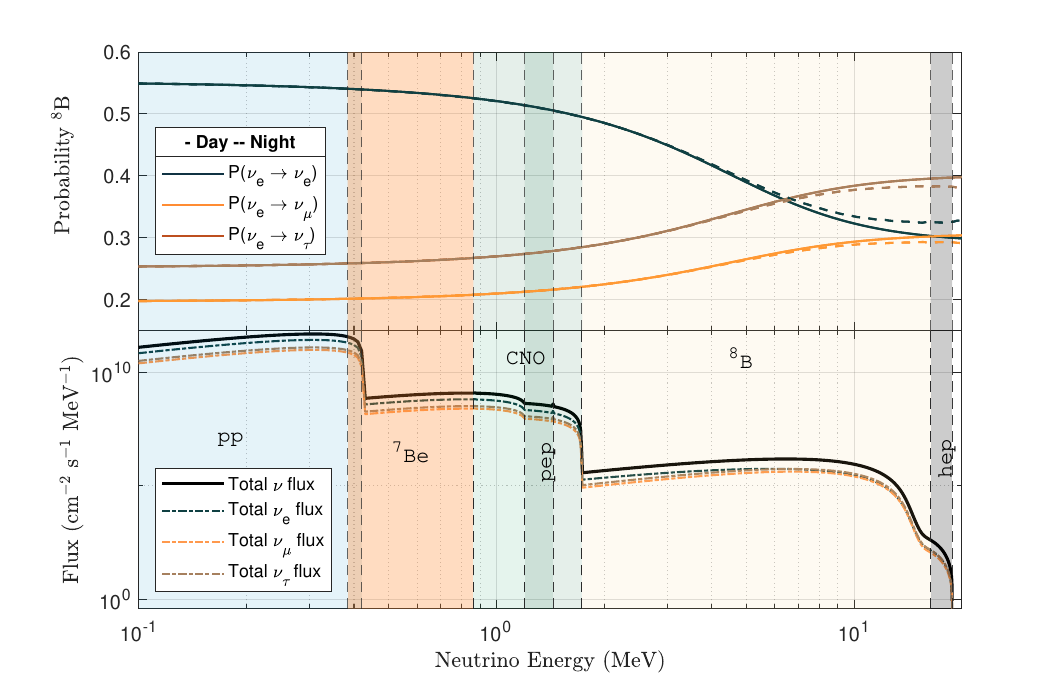}
    \caption{Survival and appearance probabilities (top) and fluxes (bottom) as a function of neutrino energy for the different components of the solar neutrino flux. The top panel shows the day and night probabilities for the $^8$B flux.  Labels for the flux components show the energy range over which each component is prominent. 
    \label{fig:flux and probability}}
\end{figure*}

 For the case of a 10 MeV neutrino, Figure~\ref{fig:zenith} shows the average values of the  probabilities using Equation~\ref{eq:pe}. Using these results, Figure~\ref{fig:flux and probability} shows the survival and appearance probabilities specifically for the $^8$B component of the flux, as a function of neutrino energy. This component is highlighted because it will be the most readily detectable through CE$\nu$NS at dark matter detectors, given the nuclear recoil thresholds of these detectors. This is shown explicilty in Table~\ref{tab:Solar neutrino flux}, which shows the threshold energy required for detection of each of the solar flux components. Figure~\ref{fig:flux and probability} shows that the appearance probability for $\nu_\tau$ is largest, $\sim 40\%$, as compared to the $\nu_\mu$ and $\nu_e$ components, which are $\sim 30\%$. Matter effects are largest for the highest energy neutrinos; matter effects slightly reduce the probability for $\nu_\tau$ and $\nu_e$ at high energy, and decrease the probability for $\nu_\mu$ at high energy. Below we exploit this difference between the day and night probabilities to determine the day-night asymmetry in the event rate. 

\section{Results} 
\label{sec:results} 
\par In this section we present our results, starting with the predicted event rates in Argon and Xenon detectors, and we then examine the prospects for detecting radiative corrections to the cross section. We then present the calculation of the day-night asymmetry, and estimate the possibilities for detection. 

\subsection{Event rates for Argon and Xenon detectors} 
 
\par The event rate for a $\nu_\alpha$ is
\begin{equation}
    \frac{dN_\alpha}{dT} = \int_{E_{\nu, min}} \frac{d\phi(E_\nu)}{d E_\nu}\frac{d\sigma_\alpha(E_\nu,T)}{dT} P(\nu_e\rightarrow\nu_\alpha) dE_\nu
    \label{rate}
\end{equation} 
where $d\phi/d E_\nu$ is the neutrino spectrum produced in the Sun. The integral bounds starts from $E_{\nu, min}$, which corresponds to the minimum energy of the neutrino to produce nuclear recoils of energy $T$, and is given by $E_{\nu, \textrm{min}} = \frac{1}{2}(T + \sqrt{T^2 + 2TM_A})$. Equation~\ref{rate} differs from the typical formula for CE$\nu$NS event rates through the inclusion of the survival and appearance probability terms $P(\nu_e\rightarrow\nu_\alpha)$. 

\par Defining a minimum detector threshold energy for nuclear recoils as $T_{th}$, the number of events for a given neutrino flavor is
\begin{equation}
    N_\alpha = \int_{T_{th}} \frac{dN_\alpha}{dT} dT. 
\end{equation}
The total event rate and number of events as seen by the detector is then 
\begin{eqnarray}
     \frac{dN}{dT} &=& \sum_\alpha \frac{dN_\alpha}{dT} = \frac{dN_e}{dT} +\frac{dN_\mu}{dT} +\frac{dN_\tau}{dT} \nonumber \\
     N_{total} &=& \sum_\alpha N_{\alpha} = N_e + N_\mu + N_\tau \nonumber
\end{eqnarray}

\par
\begin{figure*}[h]
    \centering
    \includegraphics[width=0.49\textwidth]{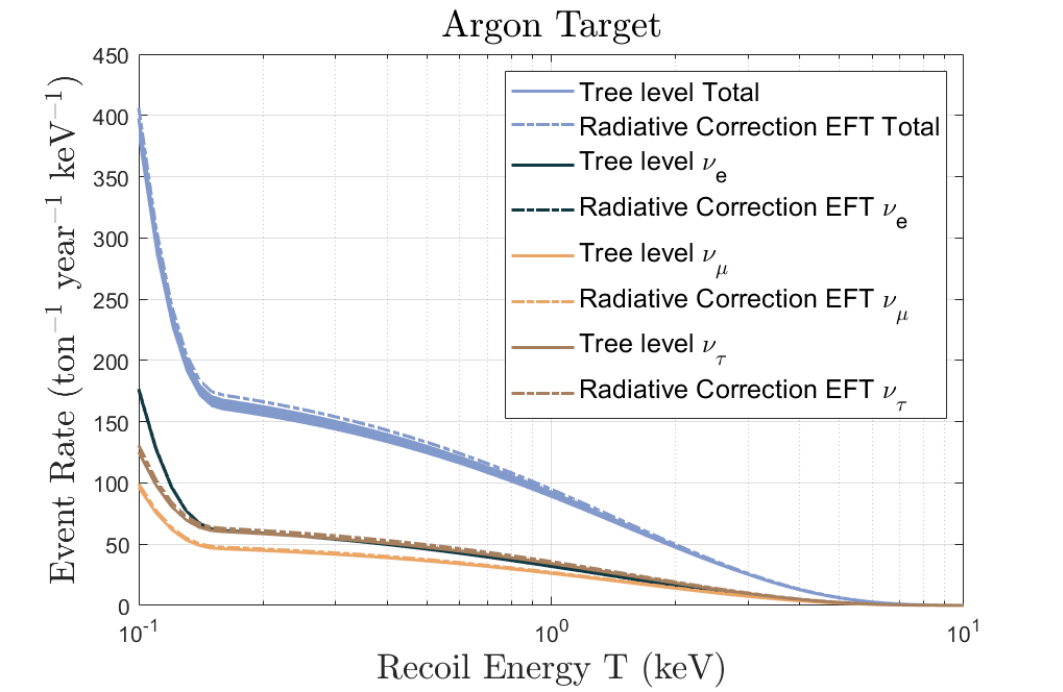}
    \includegraphics[width=0.49\textwidth]{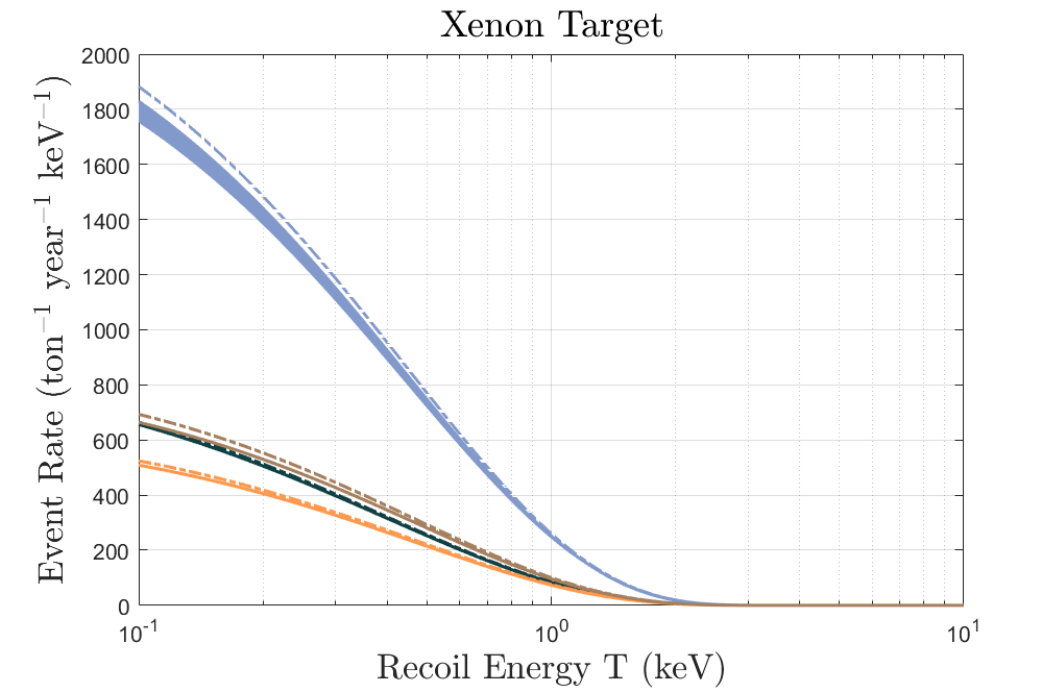}
    \caption{Event rates for Argon (left) and Xenon (right) as a function of nuclear recoil energy. Shown are the tree-level calculations as compared to the model including radiative corrections. The band for the tree level calculations show the effect of changing the value of $\sin^2 \theta_w$. }
    \label{fig:eventrate}
\end{figure*}

\begin{figure*}[h]
    \centering
    \includegraphics[width=0.49\textwidth]{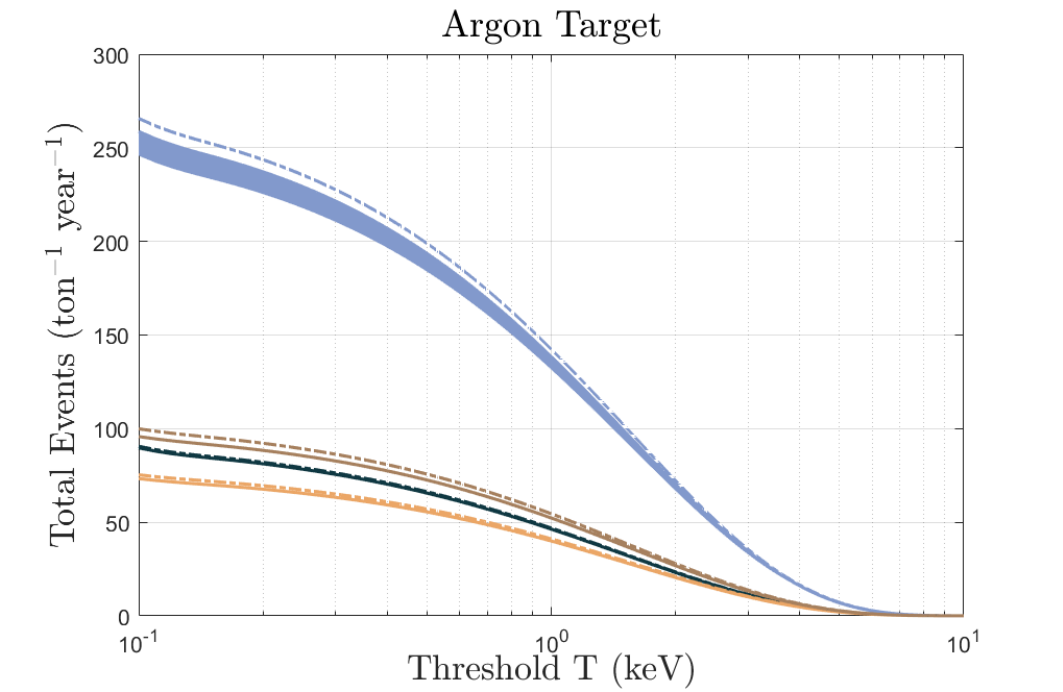}
    \includegraphics[width=0.49\textwidth]{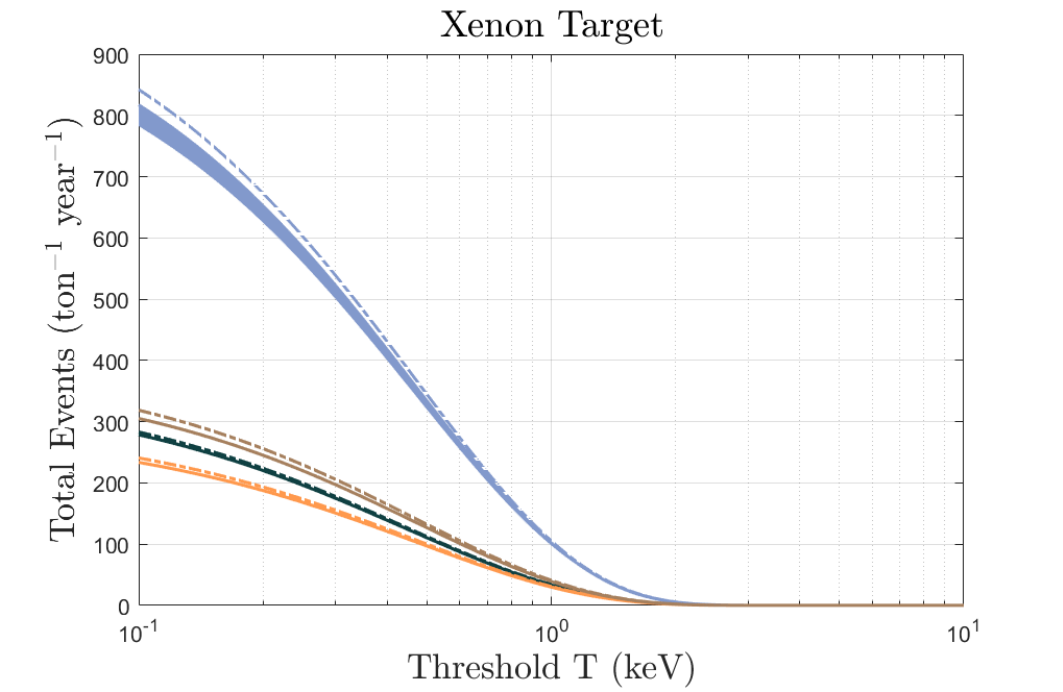}
    \caption{Total event rate above a recoil threshold energy for Argon (left) and Xenon (right) as a function of threshold nuclear recoil energy. Shown are the tree-level calculations as compared to the model including radiative corrections. The band for the tree level calculations show the effect of changing the value of $\sin^2 \theta_w$. Refer to Figure~\ref{fig:eventrate} for the legend.}
    \label{fig:totalevents}
\end{figure*}

 Figure~\ref{fig:eventrate} shows the event rate for the tree-level calculation, as compared to the rates for the radiative correction model discussed in Section~\ref{sec:crosssection}. We show the rates for Argon and Xenon detectors, which are representative targets for next generation detectors~\cite{Aalbers:2022dzr}. Figure~\ref{fig:totalevents} shows the corresponding total number of events above a threshold as a function of threshold energy. 

\par These figures also show the impact of assuming different values of $
\sin^2 \theta_w$ discussed in Section~\ref{sec:crosssection} in the tree level calculation. For both targets, we see that the corrections are largest for the $\tau$ component, and the smallest for the electron neutrino component. For both of these figures, we assume an ideal detector configuration with perfect efficiency and energy resolution. More realistically, these curves would need to be modified by the appropriate efficiency and resolution. For Argon, nearly the entire rate is due to $^8$B, with the increase at low recoil energy in Figure~\ref{fig:eventrate} from the CNO flux. On the other hand, for Xenon the rate is entirely due to $^8$B. 

\subsection{Detection prospects}\label{sec: 3 flv ellipses} 
\par We consider detectors that are only able to  measure energy deposition from the scattered nucleus, and therefore will not be able to identify scattering due to the different flavors on an event-by-event basis. Therefore, extracting out the flavor composition of the event rate must be done statistically, since we are trying to extract the fraction of the $\nu_e$, $\nu_\mu$, and $\nu_\tau$ events from a single observation, which is the number of events as a function of energy. 

\par We consider a simple model in which there are no experimental backgrounds, and assume that the detectors have perfect efficiency and energy resolution. For our mock experiment, we bin the data into a total of $n_{bin}$ recoil energy bins. Assuming a poisson distribution for the energy bins, the binned log likelihood is given by;
\begin{equation*}
    \ln{\mathcal{L}} = \sum_{i=1}^{n_{bin}} n_i \ln{\mu_i} - \mu_i - \ln{(n_i!)}
\end{equation*}
where $n_i$ refers to the number of events observed in the $i^{th}$ bin and $\mu_i$ is the central value of the expectation in the  $i^{th}$ bin. With this model, we consider two directions for the analysis. First, differentiating the tree-level from the radiative corrections model, and second, specifically isolating the $\nu_\tau$ flux component. 

\subsubsection{Differentiating Radiative-corrections from Tree-level}
In the full three-flavor model, we define the expected number of events in the $i^{th}$ bin as 
\begin{align}
    \mu_i &= \sum_{\alpha}f_{\alpha}\,\mu_{i\alpha} = f_e \mu_{ie} + f_\mu \mu_{i\mu} + f_\tau \mu_{i\tau} \nonumber\\
    &= f_e N_{ie} + f_\mu N_{i\mu} + f_\tau N_{i\tau} \nonumber\\
    &= T_{exp}( f_e \Bar{N}_{ie} + f_\mu\Bar{N}_{i\mu} + f_\tau\Bar{N}_{i\tau})
    \label{eq: mui}
\end{align}
where $\mu_{i\alpha}$ represents the expectation of $\nu_\alpha$ events in the $i^{th}$ bin, which can be taken as the number of events $N_{i\alpha}$ in that bin. In Equation~\ref{eq: mui} $T_{exp}$ is the experimental exposure, which is the size of the detector times the runtime of the experiment and $\Bar{N}_{i\alpha}$ represents the number of events for unit ton-year.

The terms $f_{\alpha}$ are equal to unity for the fiducial model, and can be interpreted as normalizations of the product of the flux and the survival probabilities for the different flavors. In other words, take $\phi_{\alpha} \sim (\phi P_{e\alpha})$ to be calculated using a fiducial set of fixed parameters, which we refer to as standard solar model (SSM) parameters. For the SSM, we use the GS98-SFII model, as shown in Table 2 of Ref.~\cite{Haxton:2012wfz}. Then $f_\alpha$ budgets for the uncertainty in this calculation, and we have 
\begin{align*}
    f_\alpha = \frac{(\phi_{\alpha})}{(\phi _{\alpha})_{SSM}}. 
\end{align*}

\par We then differentiate $\ln{\mathcal{L}}$ with respect to the model parameters $f_\alpha$, and average over the likelihood function. The results are the elements of the inverse covariance matrix, or the Fisher matrix~\cite{coe2009fisher} $F$, which are given by
\begin{align}
      F_{\alpha\beta} = \sum_{i=0}^{n_{bin}}T_{exp} \frac{\Bar{N}_{i\alpha}\Bar{N}_{i\beta}}{\Bar{N}_{itot}}. 
      \label{eq: fisher}
\end{align}
 From the inverse of the Fisher matrix, we can generate the variance on the $f_\alpha$'s, i.e. $\langle f_\alpha \rangle$, as well as the correlation and containment regions on these parameters. Priors on the $f_\alpha$'s can be included via the methods discussed in the ref~\cite{coe2009fisher}.

\par Even though we have summed over all components of the solar flux (Figure~\ref{fig:flux and probability}), for the thresholds under consideration for the CE$\nu$NS process in the detectors, the event rates are dominated by the $^8$B component of solar flux, as can be explicitly seen in Table~\ref{tab:Solar neutrino flux}. An updated global analysis finds that the $^8$B flux uncertainty is $\sim 2.5\%$~\cite{Orebi_Gann_2021}; motivated by these results we take a prior of $2.5\%$ on the $f_\alpha$'s. To additionally account for theoretical uncertainties on the components of the oscillation parameters, we add $10\%$ systematic uncertainties on the observable $\Bar{N}_{itot}$. This is intended to conservatively encapsulate the uncertainty on the parameters of the neutrino mixing matrix. Note that we do not account for theoretical uncertainties on the parameters that describe the cross section, and assume that these are fixed by the model. 

\par From the flux normalizations described above, we can define the flux-averaged cross section for a given flavor as
\begin{equation}
\langle \sigma_\alpha \rangle = \frac{\int \int \frac{d\phi_\alpha}{dE_\nu} \frac{d\sigma_\alpha}{dT} dE_\nu dT}{\int \frac{d\phi_\alpha}{dE_\nu} dE_\nu}
\label{eq:fluxaveragecrosssection}
\end{equation}
where the integral is above the threshold recoil energy and over the kinematically-accessible range of neutrino energies. Since our base parameters, the $f_\alpha$'s, above are proportional to the total number of events for each flavor component, we must transform to obtain projected errors on the flux-averaged cross section. This transformation is done by calculating the covariance matrix, $C$, as
\begin{equation}\label{eq:errors}
            C = \mathrm{J} F^{-1} \mathrm{J}^{\top},
\end{equation}
where $J$ is the Jacobian matrix given by $J_{\alpha \beta}=\frac{\partial  \langle \sigma_\beta \rangle}{\partial f_\alpha}$ and $F^{-1}$ is the inverse of the Fisher matrix given in Equation~\ref{eq: fisher}. We then determine the pair-wise confidence regions by calculating a $\chi^2$ surface:
\begin{equation}\label{eq:chi2}
            \chi^2 = (\langle \sigma_\alpha \rangle-\langle \sigma_\alpha \rangle_{\mathrm{model}})(C)^{-1}(\langle \sigma_\alpha \rangle-\langle \sigma_\alpha \rangle_{\mathrm{model}})^{T}
\end{equation}
where $\Delta \chi^2 = 2.3,6.17$ correspond to $1\sigma$ and $2\sigma$ respectively. To determine the fiducial values of $\langle \sigma_\alpha \rangle_{model}$, we use $N_{\alpha} / (T_{exp}\langle\phi_{\alpha}\rangle)$ for a given lower recoil energy threshold, where $\langle\phi_{\alpha}\rangle$ is the average flux, also in the denominator in Equation~\ref{eq:fluxaveragecrosssection}.

\par
\begin{figure*}[h]
    \hspace*{-1.3cm}
    \includegraphics[width =1.1\textwidth]{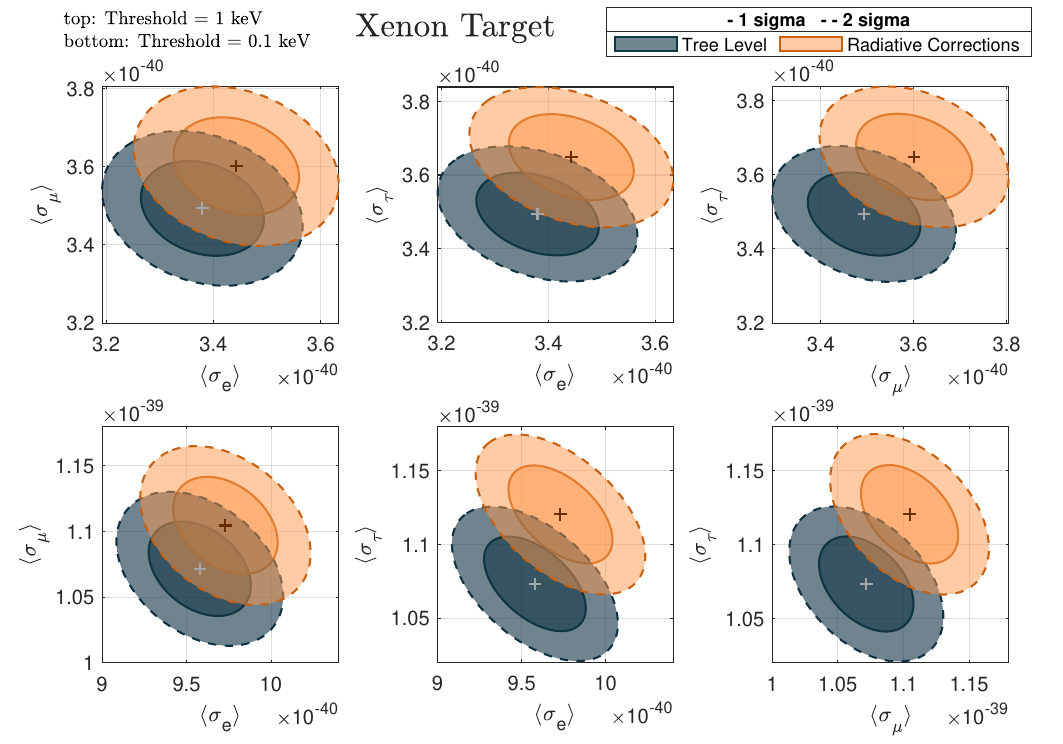} 
    \caption{One and two sigma containment regions for the flux-averaged cross section, in units cm$^2$, in the full three-flavor model assuming a Xenon target, for an exposure of 100 ton-yr. Containment regions are shown for the tree level cross section model, as well as the radiative correction model. The top row assumes a nuclear recoil threshold of 1 keV, and the bottom shows a threshold of 0.1 keV. All panels assume a flux prior on solar neutrinos of 2.5\% and $\sin^2 \theta_w = 0.23857$.}
    \label{fig:3flavor2.5perecentXe}
\end{figure*}

\begin{figure*}[h]
    \hspace*{-1.3cm}
    \includegraphics[width =1.1\textwidth]{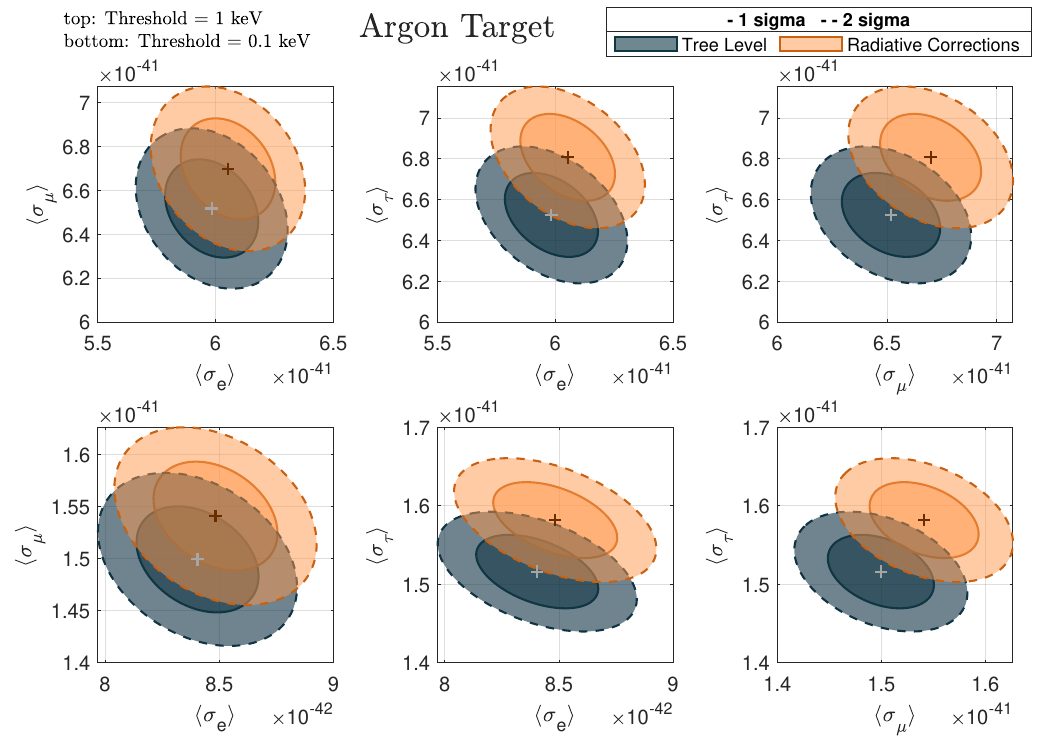}
    \caption{One and two sigma containment regions for the flux-averaged cross section, in units cm$^2$, in the full three-flavor model assuming an Argon target, for an exposure of 100 ton-yr. Containment regions are shown for the tree level cross section model, as well as the radiative correction model. The top row assumes a nuclear recoil threshold of 1 keV, and the bottom shows a threshold of 0.1 keV. All panels assume a flux prior on solar neutrinos of 2.5\% and $\sin^2 \theta_w = 0.23857$.}
    \label{fig:3flavor2.5perecentAr}
\end{figure*}

 The projected uncertainties on $\langle \sigma_\alpha \rangle$ for different threshold recoil energies and for different combinations of neutrino flavor for $T_{exp} = 100$ ton-yr are shown in Figure~\ref{fig:3flavor2.5perecentXe} and Figure~\ref{fig:3flavor2.5perecentAr} for Xenon and Argon, respectively. For $2.5\%$ prior, in the $\langle \sigma_\mu \rangle$-$\langle \sigma_\tau \rangle$ parameter space the tree-level and the radiative correction models are distinguishable for an optimistic threshold of 0.1 keV at the 1-sigma level, while the models are less distinguishable for the 1 keV threshold. 

\par The shift in the flux-averaged cross sections from the tree-level to the radiative correction are evident in all cases. Note that even though the flux averaged cross section shifts from the radiative to the tree-level calculations, this shift is similar in all flavor directions, and is primarily due to the flavor-independent contributions. The negative slope of the semi-major axis implies  the $\langle \sigma_\alpha \rangle$'s are anti-correlated.

\par
\begin{figure*}[ht]
    \hspace*{-1.3cm}
    \includegraphics[width =1.1\textwidth]{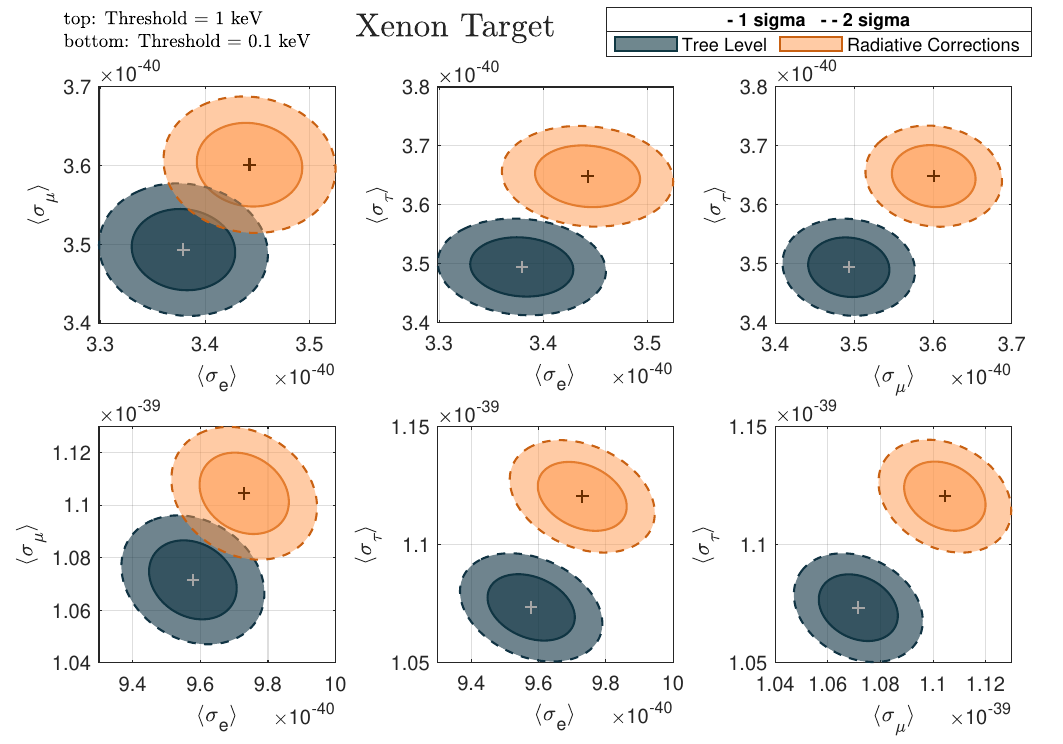}
    \caption{Same as Figure~\ref{fig:3flavor2.5perecentXe}, except assuming a flux prior of 1\% and $\sin^2 \theta_w = 0.23857$}
    \label{fig:3flavor1perecentXe}
\end{figure*}

\begin{figure*}[h]
    \hspace*{-1.3cm}
    \includegraphics[width =1.1\textwidth]{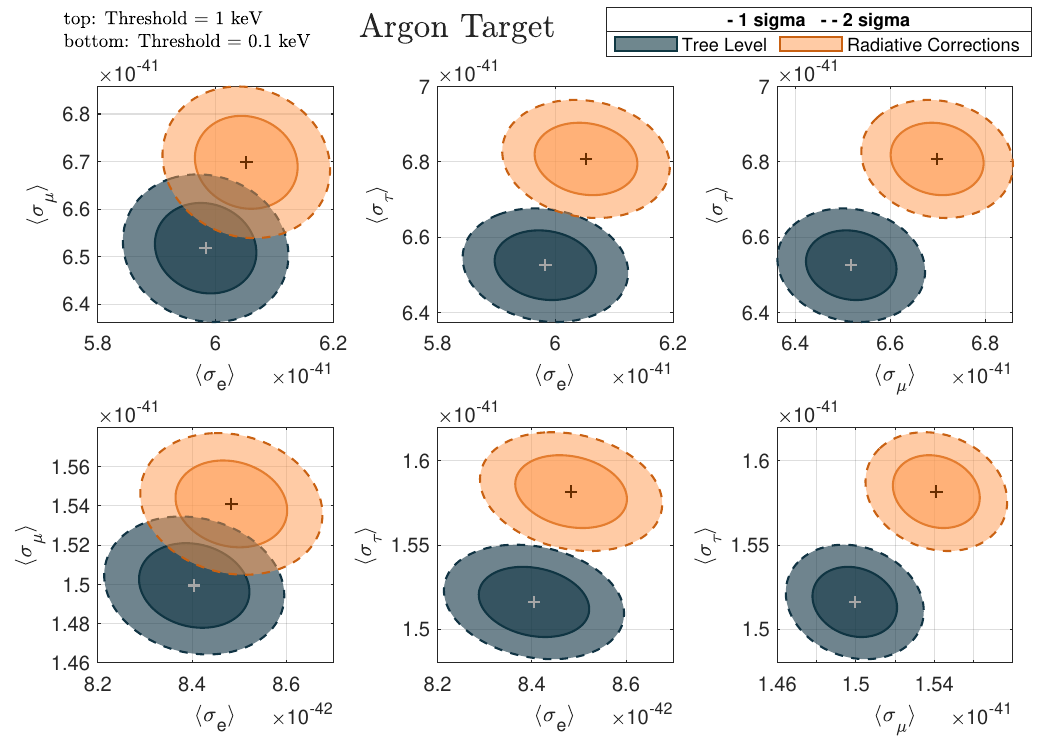}
    \caption{Same as Figure~\ref{fig:3flavor2.5perecentAr}, except assuming a flux prior of 1\% and $\sin^2 \theta_w = 0.23857$}
    \label{fig:3flavor1perecentAr}
\end{figure*}

 In the future, the precision of solar neutrino flux measurements may be improved. To show how our results improve with more precise flux measurements, Figures~\ref{fig:3flavor1perecentXe} and ~\ref{fig:3flavor1perecentAr} 
exhibit the chance of distinguishing the tree level and radiative corrections flux average cross-sections for a 1\% flux prior. In this case, especially in the $\nu_\mu$ and $\nu_\tau$ space, we see that the radiative cross section is distinguishable from the corresponding tree-level cross sections. We see that it is possible to separate the $\nu_\mu$ and $\nu_\tau$ components for both Xenon and Argon targets. This would be the first such separation of $\nu_\mu$ and $\nu_\tau$ components of the Solar neutrino flux. In the appendix, we examine the effect of changing the value of $\sin^2 \theta_w$ on the discrimination between the tree-level and radiative models. 

\subsubsection{Effective two-flavor model for detection of $\nu_\tau$}

\par In the analysis above we have presented a methodology for isolating the $\tau$ neutrino flux within the context of a full three-flavor analysis. Even though the cross section is different when including radiative corrections, the measurement is still challenging, in large part because of the degeneracy between the three flavors. Since the $\tau$ neutrino flux from the Sun has never been specifically isolated, it is interesting to approach the analysis in a simpler manner. With this motivation, we consider an effective two-flavor model, in which the muon and electron neutrino flux components are combined, so that the two model parameters are given by the tau flux averaged cross section, and the weighted muon and electron flux averaged cross section. 

\par
\begin{figure*}[h]
    \includegraphics[width =0.8\textwidth]{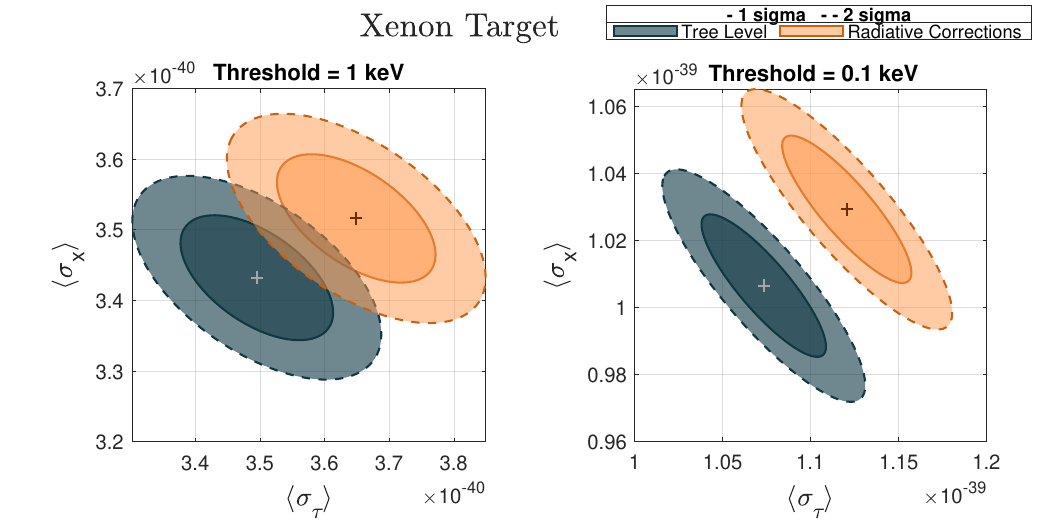}
    \caption{One and two sigma containment regions for the flux-averaged cross section, in units cm$^2$, in the effective two-flavor model assuming a Xenon target. Containment regions are shown for the tree level cross section model, as well as the radiative correction model. The left panel assumes a nuclear recoil threshold of 1 keV, and the right shows a threshold of 0.1 keV. All panels assume a flux prior on solar neutrinos of 2.5\% and $\sin^2 \theta_w = 0.23857$.}
    \label{fig:2flavor2.5perecentXe}
\end{figure*}

\begin{figure*}[h]
    \includegraphics[width=0.8\textwidth]{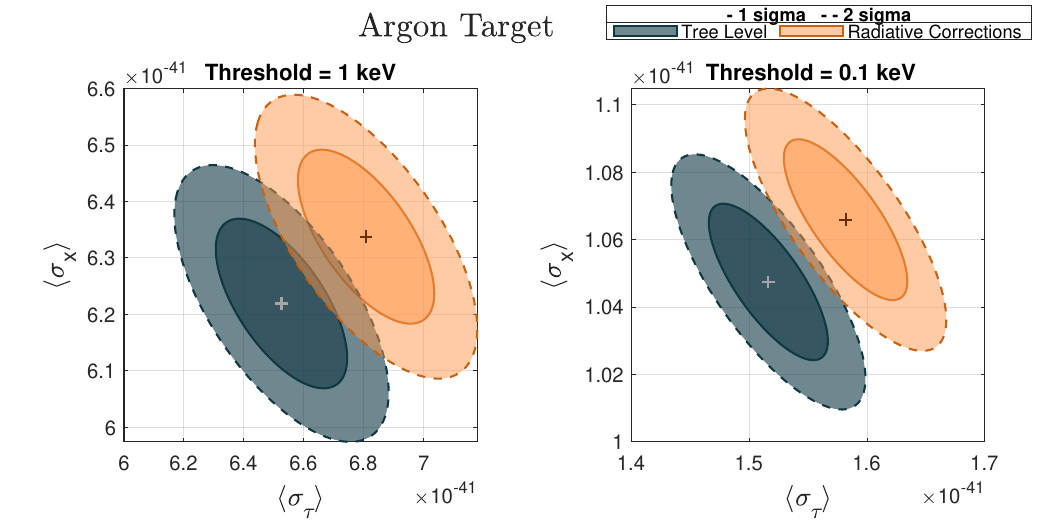}
    \caption{One and two sigma containment regions for the flux-averaged cross section, in units cm$^2$, in the effective two-flavor model assuming an Argon target. Containment regions are shown for the tree level cross section model, as well as the radiative correction model. The left panel assumes a nuclear recoil threshold of 1 keV, and the right shows a threshold of 0.1 keV. All panels assume a flux prior on solar neutrinos of 2.5\% and $\sin^2 \theta_w = 0.23857$}
    \label{fig:2flavor2.5perecentAr}
\end{figure*}

 In this effective two-flavor case, we replace Equation~\ref{eq: mui} with
\begin{align}
     \mu_i &= f_\tau N_{i\tau} + f_x(N_{i\mu} + f_\tau N_{ie}) \nonumber \\
    &= T_{exp}( f_\tau \Bar{N}_{i\tau} + f_x(\Bar{N}_{i\mu} + \Bar{N}_{ie}))
\end{align}
In this model we define the effective flux-averaged cross section for muon and electron components as 
\begin{equation} 
\langle \sigma_x \rangle = \frac{N_e + N_\mu}{T_{exp}\langle\phi_x\rangle} 
\end{equation} 
and the average flux as 
\begin{equation} 
\langle\phi_x\rangle = \langle\phi_e\rangle + 
\langle\phi_\mu\rangle.
\end{equation} 
On repeating the algorithm outlined in the previous subsection~\ref{sec: 3 flv ellipses} but in the two-parameter space comprising of $\langle \sigma_\tau \rangle$ and $\langle \sigma_x \rangle$, we obtain the projected uncertainties as shown in Figures ~\ref{fig:2flavor2.5perecentXe} and ~\ref{fig:2flavor2.5perecentAr} for $\sin^2 \theta_w = 0.23857
$ (The effect of varying $\sin^2 \theta_w$ is shown in the Appendix). For an optimistic 0.1 keV threshold, we see clear separation between the radiative and tree-level models, even at the two-sigma level for Xenon and nearly two-sigma level discrimination for Argon. Discriminiation between the radiative and tree-level models is possible at the one-sigma level for a 1 keV threshold. 

\par We note that even though in our effective two-flavor model, we have isolated the $\nu_\tau$ component, this choice is not unique; a similar analysis could have been performed isolating the other two flavors. Further, though the results presented in the effective two-flavor analysis are more aggressive than the three-flavor model, information is lost when combining the fluxes. Nonetheless, the effective two-flavor formalism does provide an important check on the full three-flavor model, and an approximate method to isolate each of the flux components. 

\subsection{Including detector efficiency and energy resolution}

In the analysis to this point, we have assumed an ideal detector. To estimate how the detection prospects change when assuming a more realistic detector configuration, including such effects as modeling the detector energy resolution and efficiency as a function of energy, we use results from the NEST simulation~\cite{Szydagis:2011tk}. For details on the parameters we use for the NEST simulation, we refer to previous related analyses~\cite{Newstead:2020fie,Zhuang:2023dzd}. 

\begin{figure*}
    \centering
    \includegraphics[width =1\textwidth]{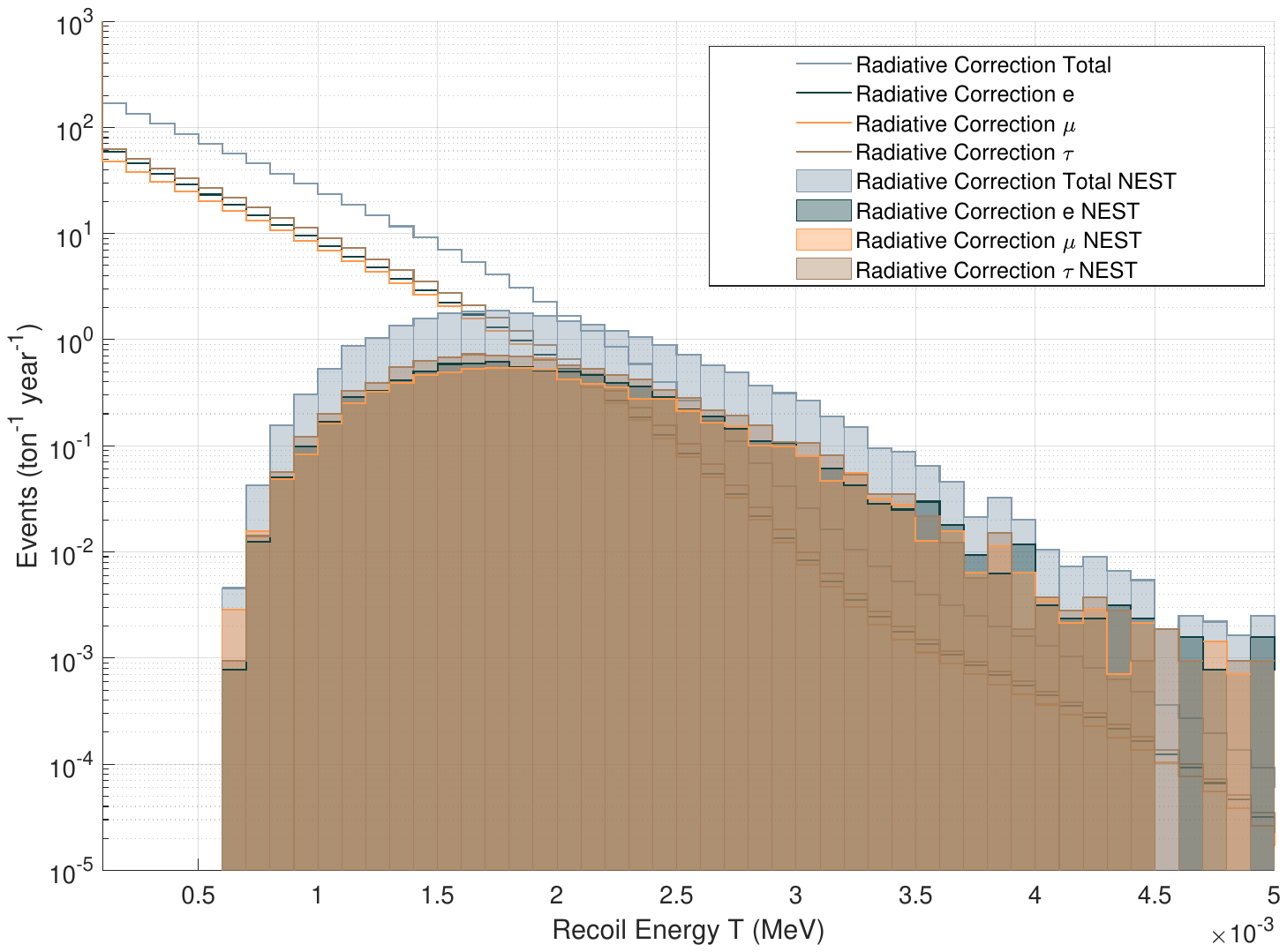}
    \caption{Event rate spectrum for the ideal case of perfect energy resolution and efficiency, as well as the corresponding spectra accounting for energy resolution and efficiency using the NEST simulation. Shown are the spectra for each of the flavors, along with the total spectrum which is the sum of all flavors.}
    \label{fig:NESTspectrum}
\end{figure*}

Figure~\ref{fig:NESTspectrum} shows the event rate spectrum obtained from NEST, compared to the ideal case for a Xenon detector. Shown are the spectra for each flavor, as well as for the sum of all flavors. This analysis shows that the efficiency drops to zero at approximately $0.5$ keV. In addition, energy resolution smears the rate so that events appear beyond the nominal endpoint of the spectrum.

\par
\begin{figure*}[h]
     \hspace*{-1.3cm}
    \includegraphics[width =1.1\textwidth]{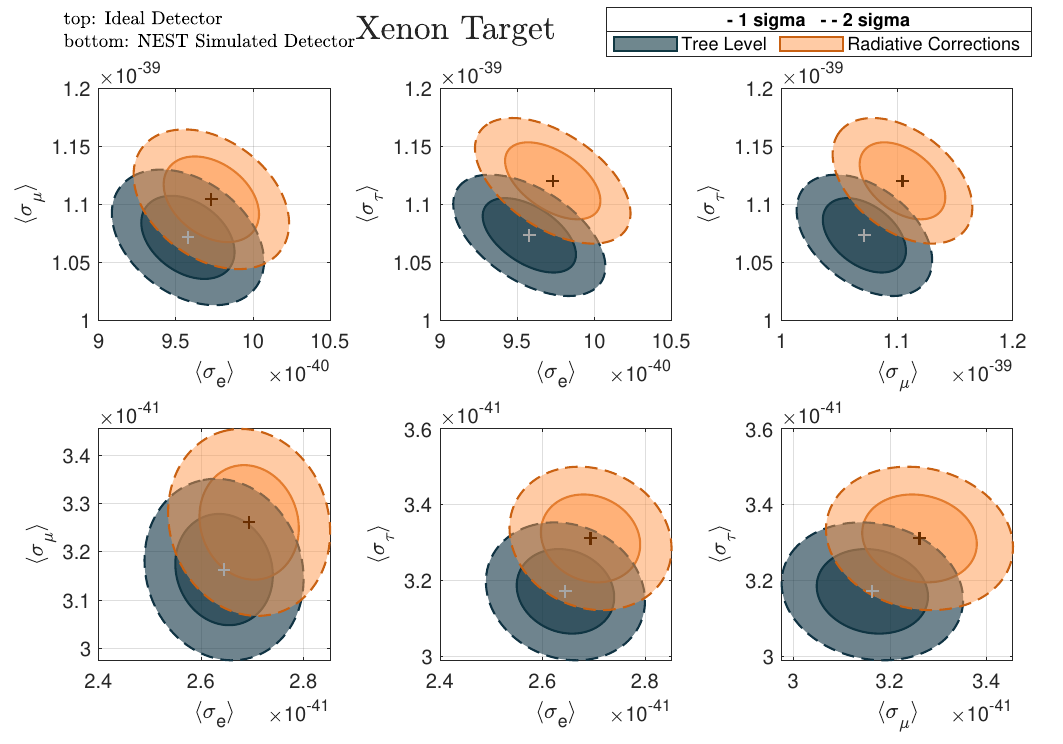}
    \caption{One and two sigma containment regions for the flux-averaged cross section, in units cm$^2$, in the full three-flavor model assuming a Xenon target. Containment regions are shown for the tree level cross section model, as well as the radiative correction model. The top row assumes ideal detector, and the bottom shows NEST simulated detector. All panels assume a flux prior on solar neutrinos of 2.5\% and $\sin^2 \theta_w = 0.23857$ for a 100 ton-year exposure.}
    \label{fig:NEST ellipses}
\end{figure*}

 Using the results from~\ref{fig:NESTspectrum}, the lower panel in figure~\ref{fig:NEST ellipses} shows the error ellipses based on the simulated data for a Xenon target. Comparing it with the ideal case shown in top panel of the same figure~\ref{fig:3flavor2.5perecentXe} we see that the containment regions have slightly larger overlaps. However, one can distinguish between the two models if one has a better prior, from better flux measurements and/or longer exposure. Additionally, if the true value of $\sin^2 \theta_w < 0.23857$, the difference between the tree level and radiative correction will become more evident. A more rigorous likelihood analysis with real detector based simulation, like above, and with other targets is left for future work.

\subsection{Day-night effect}
\par Though as discussed the flavor-dependent radiative corrections are small, since it is very interesting to search for flavor dependencies in the CE$\nu$NS cross section, we now discuss the possibilities for extracting the flavor dependencies. We will specifically consider the differences in the event rates for neutrinos that pass through the Earth as opposed to those that reach the detector without going through the Earth. This day-night effect has been detected at a level $\sim 3\%$ for $^8$B neutrinos using the charged-current channel at Super-Kamiokande~\cite{Super-Kamiokande:2016yck}. Here, we provide the first estimate of this effect using CE$\nu$NS with flavor-dependent corrections. 

\par If the number of events in the night and day are, respectively, $N_N$ and $N_D$, the asymmetry between the day night events is 
\begin{equation}
    A = 2 \frac{N_N - N_D}{N_N + N_D} 
    \label{eq:asymmetry} 
\end{equation}
For the cross sections above, we find that the day-night asymmetry is 
\begin{align}
A \simeq -2.5\sim 10^{-4}
    \label{eq:A} 
\end{align} 
at the thresholds under consideration. The asymmetry is the strongest at the highest recoil energies as can be seen in the top panel of Figure;~\ref{fig:flux and probability}, though at these energies the event rates are the lowest~\ref{fig:totalevents}. The above asymmetry is nearly two orders of magnitude less than the asymmetry at Super-Kamiokande, and about an order of magnitude less than the estimation of the asymmetry that may be detectable by Borexino~\cite{Ioannisian:2015qwa,BOREXINO:2022wuy}. 

\par We now examine the prospects for detection of the small asymmetry in Equation~\ref{eq:A}. Making the approximation $N_N \simeq N_D$, we have 
\begin{equation}
    A N_N \simeq N_N - N_D 
\end{equation}
The uncertainty on the event rate is 
\begin{equation}
    \sigma  = \sqrt{ 2 N_N } 
\end{equation}
So to get an $x$ sigma detection, we have
\begin{equation}
    x = \frac{N_N - N_D}{\sqrt{2 N_N } } \simeq \frac{A N_N}{\sqrt{2 N_N}} = \frac{A}{\sqrt{2}} \sqrt{N_N}  
\end{equation}
From this estimate, we see that accumulating the largest number of events is ideal, which means reducing the threshold to the lowest values possible. However, the matter effects are most significant at the highest recoil energies, so in order to most easily detect the effect, it is prudent to focus on the highest recoil energies, where the flux is the smallest. Integrating down to 0.1 keV in Xenon, the rough estimate above implies that exposures of $\mathcal{O}(10^5)$ ton-yr are required to detect the asymmetry at 1-$\sigma$, which are larger than currently planned detectors.
 
\section{Discussion and conclusion} 
\label{sec:discussion}
\par We have discussed the prospects for identifying radiative corrections to the CE$\nu$NS cross section using the solar neutrino flux at next generation dark matter detectors. Solar neutrinos are an interesting source because at neutrino energies $\sim 1-10$ MeV, the nuclear form factor effects are small, so the coherence condition is well-preserved. Further, the fluxes are measured to high precision by combining results from previous experiments. Within the context of a full three-flavor analysis that includes the effects of matter oscillations in the Sun and the Earth, we find that detectors with exposure $\sim 100$ ton-year would be able to measure a cross section value that deviates from the tree-level prediction. 

\par Because the CE$\nu$NS cross section including radiative corrections is different for each of the three flavors, our analysis provides a novel means to study the transitions to mu and tau neutrino flavors. This would provide a means to measure the tau and muon neutrino fluxes by exploiting both their differences in fluxes and cross sections. More generally, it would provide a novel method to study tau neutrino interactions at low energy, and to study tau neutrinos using CE$\nu$NS.

\par Furthermore, it is interesting to consider if the separation of muon and tau neutrino fluxes may be possible with existing neutral current solar neutrino data from, for example, SNO or Super-Kamiokande. However, this would only rely on the differences between the muon and tau fluxes. This would represent a complementary extension to our analysis.

\par For the first time, we have provided an estimate of the day-night asymmetry in the $^8$B scattering rate with flavor-dependent radiative corrections included. We find that flavor-dependent corrections induce a small day-night asymmetry of $\sim - 3 \times 10^{-4}$ in the event rate. Though it would provide a novel probe of flavor oscillations, this appears to be challenging to detect even with next generation dark matter detectors.

\par While in our analysis we have focused on the detection of radiative corrections using $^8$B neutrinos through the CE$\nu$NS channel, our results can be extended in the future to other channels, such as the neutrino-electron scattering channel. In this case, the pp component of the solar flux induces the largest number of events, which is expected to be measured to high precision in future experiments. Similar to CE$\nu$NS, radiative corrections induce a few percent shift in the cross section from the tree-level value~\cite{Tomalak:2019ibg}, to which future experiments should be sensitive. We leave this topic for future study.

\par For our primary analysis, we have assumed idealized configurations for the detectors. In particular, our results have assumed energy thresholds that are at this time lower than the scale at which current detectors are operating. However, with possible improvements in detector technology and analysis, measurements such as those we consider may be achievable. We have provided an estimate for the corrections that arise due to finite detector energy resolution and efficiency, and find that the constraints in this case are slightly weakened. Though even in this case of finite resolution and efficiency, radiative corrections would be detectable with several hundreds of ton-years of exposure. Our simulation results may be improved upon by running more realistic simulations of the neutrino interactions in Argon or Xenon~\cite{2011NEST}.

\par Additional interesting phenomenology in the neutrino sector may also be considered using the results of our analysis. For example, we may consider how changes in the CP-violating phase change the three-flavor transition probabilities in the mu-tau sector. We leave these and related interesting questions to future work.

\begin{acknowledgments}
L.S. and N.M. are supported by the DOE Grant No. DE-SC0010813. We are very grateful to O. Tomalak and R. Plestid for several discussions on radiative corrections and CE$\nu$NS, and K. Kelly  and  P. Machado for the discussion of the formalism of neutrino oscillations. We thank Yi Zhuang for providing the spectra output from the NEST simulations. 
\end{acknowledgments}

\bibliography{apssamp}

\appendix 
\section{Effect of change in $\sin^2 \theta_w$ }

\begin{figure*}[h]
    \hspace*{-1.3cm}
    \includegraphics[width =1.1\textwidth]{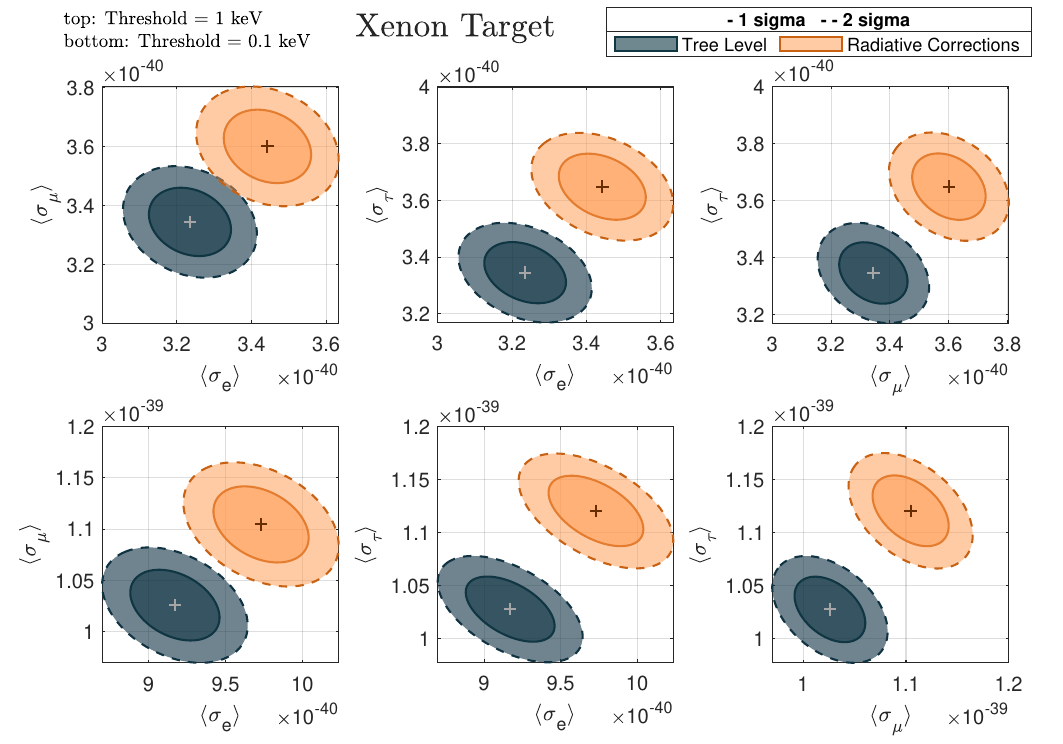 }
    \caption{Same as Figure~\ref{fig:3flavor2.5perecentXe}, except assuming $\sin^2 \theta_w$ = 0.23112}
    \label{fig:3flavor2.5perecentXe sin_w=0.23112}
\end{figure*}

\begin{figure*}[h]
    \includegraphics[width =0.8\textwidth]{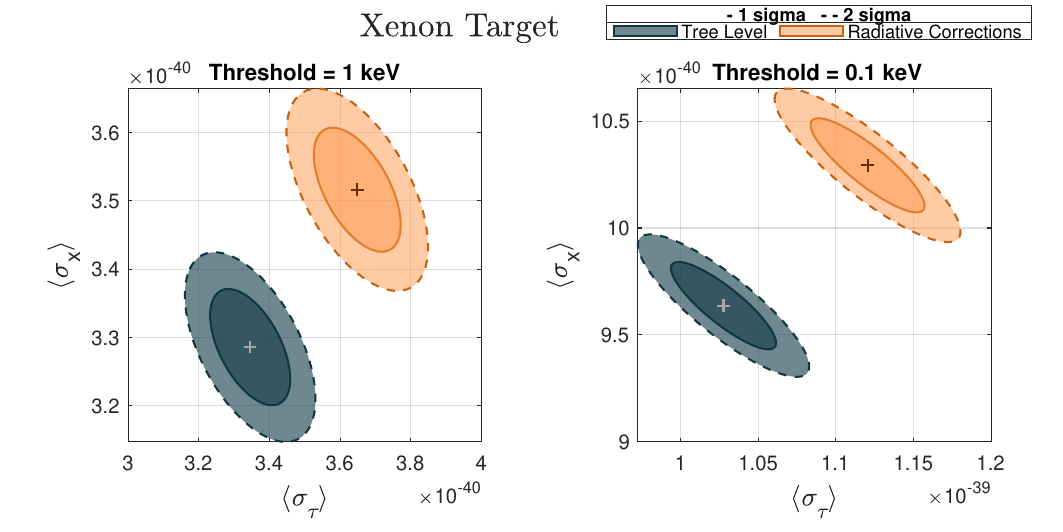}
    \caption{Same as \ref{fig:2flavor2.5perecentXe}, except assuming $\sin^2 \theta_w$ = 0.23112}
    \label{fig:2flavor2.5perecentXe sin_w=0.23112}
\end{figure*}

\begin{figure*}[h]
    \hspace*{-1.3cm}
    \includegraphics[width =1.1\textwidth]{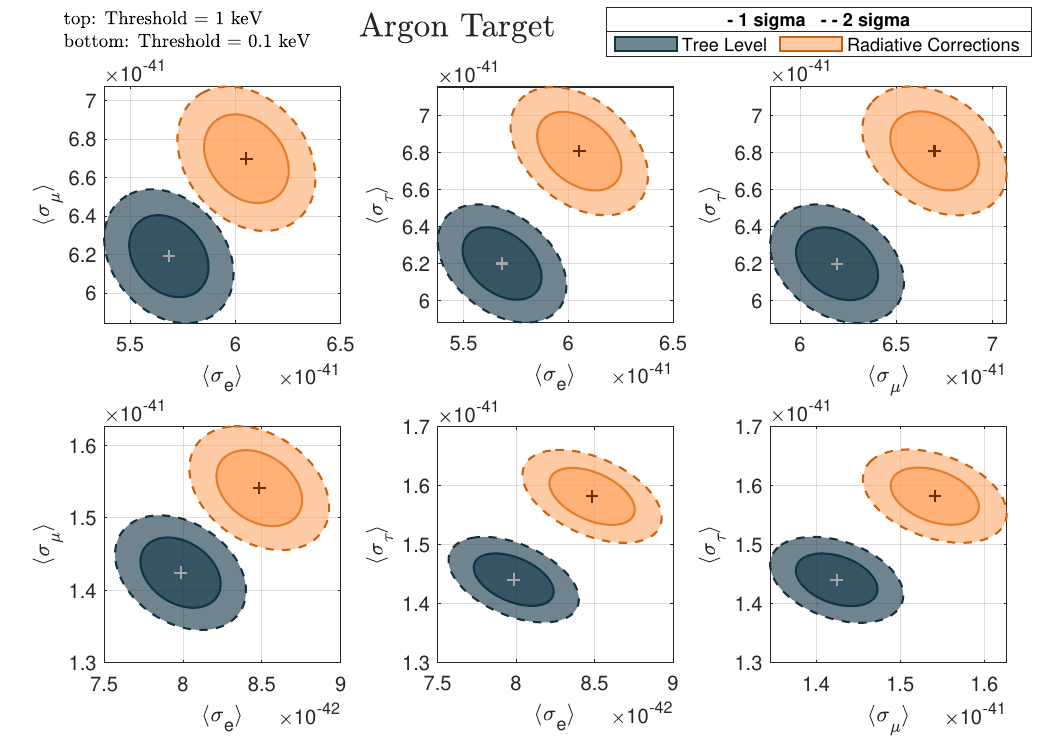}
    \caption{Same as Figure~\ref{fig:3flavor2.5perecentAr}, except assuming $\sin^2 \theta_w$ = 0.23112}
    \label{fig:3flavor2.5perecentAr sin_w=0.23112}
\end{figure*}

\begin{figure*}[h]
    \includegraphics[width =0.8\textwidth]{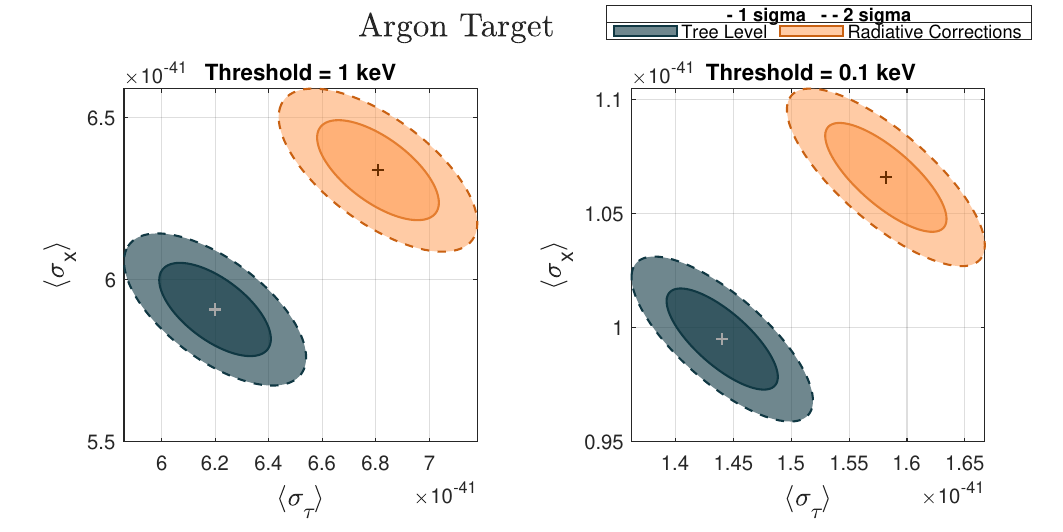}
    \caption{Same as Figure~\ref{fig:2flavor2.5perecentAr}, except assuming a 2.5\% flux prior and $\sin^2 \theta_w$ = 0.23112}
    \label{fig:2flavor2.5perecentAr sin_w = 0.23112}
\end{figure*}

In this appendix, we examine the effect of changing $\sin^2 \theta_w$ in our analysis. For an optimistic projection of $\sin^2 \theta_w = 0.23112$, the percentage difference between the radiative corrections and tree level value of the cross-section is $\sim 8\%$. This leads to a better chance of distinguishing the tree-level error models from the ones with radiative correction in both Xenon and Argon targets, as shown in Figures~\ref{fig:3flavor2.5perecentXe sin_w=0.23112},~\ref{fig:2flavor2.5perecentXe sin_w=0.23112},\ref{fig:3flavor2.5perecentAr sin_w=0.23112}, and ~\ref{fig:2flavor2.5perecentAr sin_w = 0.23112}. The strong sensitivity to $\sin^2 \theta_w$ highlights the importance of even more precise measurements of this parameters in the future~\cite{akimov2022coherent}. 

\end{document}